\documentclass[a4paper,fleqn]{cas-sc}
\usepackage[numbers]{natbib}
\usepackage {graphicx}
\usepackage{color}
\usepackage{subfigure}
\usepackage[]{hyperref}
\hypersetup{
    colorlinks = true,
    linkcolor = blue,
    anchorcolor = blue,
    citecolor = blue,
    filecolor = blue,
    urlcolor = blue
}
\usepackage{wrapfig}
\usepackage{amssymb}
\usepackage{amsthm}
\usepackage{mathtools}  

\newtheorem{proposition}{Proposition}

\newcommand{\jgrv}[1]{{\color{black} #1}}

\newcommand{\carrew} {\hfill $\Box$}
\newcommand{\carre}{\begin{flushright} \rule{2mm}{2mm} \end{flushright}}

\begin{document}
    \let\WriteBookmarks\relax
    \def\floatpagepagefraction{1}
    \def\textpagefraction{.001}
    \shorttitle{ Matching Equations of Kinetic Energy Shaping }
    \shortauthors{M. Reza J. Harandi et~al.}
    \title [mode = title]{On the Matching Equations of Kinetic Energy Shaping in IDA-PBC}


    \author[mymainaddress]{M.~Reza~J.~Harandi}
    \ead{jafari@email.kntu.ac.ir}
    \author[mymainaddress]{Hamid~D.~Taghirad$^*$}
    \cortext[cor1]{Corresponding author}
    \ead{taghirad@kntu.ac.ir}

    \address[mymainaddress]{Advanced Robotics and Automated Systems
        (ARAS), Faculty of Electrical Engineering,
        K. N. Toosi University of Technology, Tehran, Iran}

\begin{abstract}
Interconnection and damping assignment passivity-based control
scheme has been used to stabilize many physical systems such as
underactuated mechanical systems through total energy shaping. In
this method, some partial differential equations (PDEs) arisen by
kinetic and potential energy shaping, shall be solved analytically.
Finding a suitable desired inertia matrix as the solution of nonlinear PDEs related to kinetic energy shaping is a challenging problem.
 In this paper, a systematic approach to solve \jgrv{this matching equation} for
systems with one degree of underactuation is proposed. A special
structure for desired inertia matrix is proposed to simplify the
solution of the corresponding PDE. It is shown that the proposed method is
more general than that of some reported methods in the literature.
In order to derive a suitable desired inertia matrix, a necessary
condition is also derived. The proposed method is applied to three
examples, including VTOL aircraft, pendubot and 2D SpiderCrane
system.
\end{abstract}
\maketitle

\section{Introduction}\label{s1}
Different \jgrv{methods} to solve Partial differential equations (PDEs) with
and without boundary condition \jgrv{are} proposed in the literature. PDEs
are very common in different fields of engineering such as
thermodynamic, chemical process, wave analysis, etc. In some recently
developed control methodologies, the process of controller design is
reduced to finding the solution of such equations. Interconnection
and damping assignment passivity-based control is one of these
methods that is based on the general solution of some
PDEs~\cite{franco2019ida}. Providing new methodologies to find the
required solution of PDEs opens new horizon to design novel
controllers for many applications.

Passivity-based control (PBC) is a well-known methodology which was
first introduced in~\cite{ortega1989adaptive} to define a controller
design for stabilization through passivity. In this method, the
control objective is to stabilize the desired equilibrium point of
the system which is the minimum of a preselected storage function.
This method, which clearly reminiscent of standard Lyapunov
procedure, is  successfully applied to simple mechanical
systems that can be stabilized by shaping merely the potential
energy~\cite{ortega2001putting}. For applications that requires
kinetic energy shaping, PBC may be used \cite{ortega2013passivity},
but the structure of the system in closed--loop will be changed and
the storage function of the passive map does not have the
interpretation of total energy anymore, while an unnatural stable
invertibility requirement is imposed to the system
\cite{ortega2002interconnection}.

In order to conquer this drawback and expand the range of
applications of PBC, an extended version of this method is proposed. In this version, a storage function is not
fixed at first, but the desired structure of the closed-loop system
such as port-controlled Hamiltonian (PCH) or Lagrangian, is
selected. Then all assignable energy functions which are applicable
to this structure are obtained via the solution of  partial
differential equations. The most popular examples of this method are  the interconnection and damping assignment
(IDA)~\cite{ortega2002interconnection} and
the controlled Lagrangian~\cite{bloch2000controlled}. 
 IDA-PBC reform the
closed-loop system to a Hamiltonian structure with three matrices
containing interconnection between subsystems, damping term and
kernel of input matrix. One of the most important advantages of port
Hamiltonian modeling is that it is based on energy exchange and
dissipation of the system, thus passive structure of the system is
visible. Readers are referred to~\cite{ortega2004interconnection}
and  references therein for examples and other features of IDA-PBC.

The most difficulty of IDA-PBC, which restricts the application of
this method, \jgrv{is solving a set of PDEs called matching equations.}
Especially, in the case of underactuated robots; which have fewer
actuators than the system degrees of freedom (DOF); a nonlinear PDE
arises for kinetic energy shaping. In order to obviate this
difficulty, some solutions are reported in the literature. As
representatives, consider~\cite{gomez2001stabilization} that focuses
on robots with one degree of underactuation, where the inertia
matrix depends only on unactuated configuration. In the proposed
method the PDEs are reduced to a simple set of nonlinear ODEs that
are only solvable for some 2~DOF robots. a Similar method for PDE of
kinetic energy is reported in~\cite{
ortega2002stabilization}.
In~\cite{acosta2005interconnection} a method for transforming PDEs
to ODEs is proposed. This method is based on some restrictive
assumptions that reduces its application to few simple cases. Shaping the energy of port Hamiltonian
systems without solving PDEs is proposed in~\cite{
donaire2015shaping} for a special case of systems. Based on the results of this paper, in \cite{romero2016energy,romero2018global} energy shaping via PID controller is designed. The major disadvantage of last three mentioned papers is that they are merely applicable to the systems which satisfy some restrictive assumptions.
  A
method to simplify the PDEs associated with the potential energy for
 a class of underactuated mechanical systems is developed
in~\cite{ryalat2016simplified}. In~\cite{viola2007total}
simplification of kinetic energy PDE via change of coordinate is
analyzed. By considering these articles, one may argue
that proposing a general method which solves the matching equations
of any system, is a prohibitive task. Hence, new techniques with a
large domain of applicability are required to simplify the matching
equations especially, the PDE of kinetic energy shaping.

In this paper, a method to solve PDE of kinetic energy of mechanical
systems with one degree of underactuation is proposed. The
interconnection matrix is used as a free parameter in IDA-PBC to
transform the PDE with respect to the desired inertia matrix to some algebraic equations and a PDE with respect
to the unactuated coordinate. For this purpose, a special structure
for the desired inertia matrix is proposed to simplify the resulting
matching equation. \jgrv{By this means, the inverse of desired inertia matrix is designed such that merely the diagonal and the row and column corresponding to unactuated joint are nonzero. Note that one of the reason of difficulty of this PDE is that the desired inertia matrix is multipling into its partial difference. Hence, only one of the elements of this matrix is state-dependent, results in simplification of PDE. After this, and by suitably defining the free sub-block of interconnection matrix, the closed form of algebraic equations together with single PDE is derived. Three different cases are considered for the resulting PDE in which in two cases the solution may be found easily. We may invoke \cite{harandi2020solution} to solve the PDE in the other case.}
 A necessary condition is also proposed to ensure
suitably selection of the desired inertial matrix, since this matrix
plays a crucial role in the stabilization of an unstable equilibrium
point. In other words, an inertia matrix should be designed which
leads to a desired potential energy with a positive Hessian matrix
at the desired equilibrium point.

The paper is organized as follows. Section~\ref{s2} presents
background of IDA-PBC method for underactuated mechanical systems. A
necessary condition and an algorithm for simplifying the PDE of
kinetic energy are proposed in section~\ref{s3}. Three examples are
represented in Section~\ref{s4}. Finally, concluding remarks and
future works are summarized in Section~\ref{s5}.

\textbf{Notation}: $I_n$ denotes
$n\times n$ identity matrix, $0_{m\times n}$ is $m\times n$ zero matrix and $0_n$
is a $n$ dimensional column vector of zeros.  $x^{(i)}$ and $\xi^{(ij)}$ with
$x\in\mathbb{R}^n,\xi\in\mathbb{R}^{m\times n}$ denote $i$-th and $(i,j)$-th element
of $x$ and $\xi$, respectively.
 $e_i\in\mathbb{R}^n$ with $i\in\bar{n}$ is the Euclidean basis vector where $\bar{n}=\{1,..,n\}$. Gradient of a scalar function $f(x)$ with $x\in\mathtt{R}^n$ which is denoted by $\nabla f$  is a column vector as $\nabla f=[\frac{\partial f(x)}{\partial x^{(1)}},...,\frac{\partial f(x)}{\partial x^{(n)}}]^T$.

\section{Review of IDA-PBC methodology for simple mechanical systems}\label{s2}
In here, IDA-PBC method for underactuated mechanical systems is
reviewed.  The readers are referred to
\cite{ortega2002stabilization, acosta2005interconnection} for more
details. If it is assumed that the system has no natural damping,
the equations of motion may be written in the PCH form as
\begin{equation}
\label{1}
\begin{bmatrix}
\dot{q} \\ \dot{p}
\end{bmatrix}
=\begin{bmatrix}
0_{n\times n} & I_n \\ -I_n & 0_{n\times n}
\end{bmatrix}
\begin{bmatrix}
\nabla_q H \\ \nabla_p H
\end{bmatrix}
+\begin{bmatrix}
0_{n\times m} \\ G(q)
\end{bmatrix}
u,
\end{equation}
where  $H(q,p)=1/2p^TM^{-1}(q)p+V(q)$ is total energy of the system,
$q,p\in R^n$ are generalized position and momenta, respectively,
$M^T(q)=M(q)>0$ is the inertia matrix, $V(q)$ is the potential
energy and rank of $G(q)$ is equal to $m<n$. Suppose that the
desired structure for $H_d$ is given as follows
\begin{equation*}
H_d(q,p)=1/2p^TM_d^{-1}(q)p+V_d(q),
\end{equation*}
where $M_d(q)$ and $V_d(q)$ represent the desired inertia matrix and
potential energy function, respectively, and it is required that the
desired equilibrium point $q_*$ satisfies $q_*=\text{arg min}
V_d(q)$. The desired interconnection matrix is also given as follows
\begin{equation*}
J_d(q,p)=
\begin{bmatrix}
0_{n\times n} & M^{-1}(q)M_d(q) \\ -M_d(q)M^{-1}(q) & J_2(q,p)
\end{bmatrix}
\end{equation*}
in which the skew-symmetric matrix $J_2(q,p)$ is a free design
parameter. It is possible to split the control into
$u=u_{es}(q,p)+u_{di}(q,p)$, in which
\begin{equation}
\label{3}
\begin{split}
&u_{es}=(G^TG)^{-1}G^T\big(\nabla_q H-M_dM^{-1}\nabla_q H_d+J_2M_d^{-1}p\big)\\
&u_{di}=-K_vG^T\nabla_p H_d
\end{split}
\end{equation}
with $K_v>0$. This restricts the desired damping matrix to have the
form of
\begin{equation*}
R_d(q)=
\begin{bmatrix}
0_{n\times n} & 0_{n\times n} \\ 0_{n\times n} & GK_vG^T
\end{bmatrix}
\end{equation*}
The closed-loop system  takes the Hamiltonian form
\begin{equation}
\label{5}
\begin{bmatrix}
\dot{q} \\ \dot{p}
\end{bmatrix}
=
\begin{bmatrix}
0_{n\times n} & M^{-1}M_d \\ -M_dM^{-1} & J_2-GK_vG^T
\end{bmatrix}
\begin{bmatrix}
\nabla_q H_d \\ \nabla_p H_d
\end{bmatrix}
\end{equation}
the matching equations of the IDA-PBC can be separated into the
terms that depend on the kinetic  and the potential energies, i.e.
the terms depend on $p$ and terms which are independent of $p$,
respectively. This leads to
\begin{subequations} \label{67}
    \begin{align}
    &G^\bot (q)\{\nabla_q \big(p^TM^{-1}(q)p\big)-M_dM^{-1}(q)\nabla_q \big(p^TM_d^{-1}(q)p\big)+2J_{2}M_d^{-1}p\}=0_s,\label{6}\\
    &G^\bot (q)\{\nabla_q V(q)-M_dM^{-1}\nabla_q V_d(q)\}=0_s,\label{7}
    \end{align}
\end{subequations}
where $G^\bot\in\mathbb{R}^{s\times n}$ is left annihilator of $G$
and $s=n-m$. Equation (\ref{6}) is a nonlinear PDE respect to
positive definite desired inertia matrix. Given $M_d$, equation
(\ref{7}) is a linear PDE with respect to the desired potential
energy. Therefore, the main difficulty of these PDEs is finding
analytical solution for equation (\ref{6}).

In the sequel, we focus on the PDE of kinetic energy. Note that the
proposed method works for the system with one degree of
underactuation. The aim is to solve PDE (\ref{6}) or propose a
methodology to simplify it. Invoking~\cite{crasta2015matching}, and
by considering a special form for $M_d$ and utilizing $J_2(q,p)$,
this PDE is transformed to some algebraic equations and a single PDE
in which the unknown parameter is the unactuated diagonal parameter
of $M_d^{-1}$. Notice that $M_d$ has a critical role to ensure
$q_*=\text{arg min} V_d(q)$. Regardless of the most previous
researches such as \cite{viola2007total,donaire2016simultaneous} that just focus on solving
equation (\ref{6}) without directly considering PDE (\ref{7}), Here
a necessary condition is proposed to restrict selection of $M_d$ to
conduce a suitable $V_d$.

\section{Main results}\label{s3}
In this section a constructive method with respect to PDE of
kinetic energy is proposed. To accomplish that, \jgrv{let us introduce} a condition
on the selection of $M_d$ as stated in the following proposition.
\begin{proposition}\label{pr1}\normalfont
    Consider PDE (\ref{7}) and assume that $n-m=1$. If Hessian matrix $ \left.\frac{\partial^2 V_d}{\partial q^2}\right|_{q=q^*}$ is positive definite, then the following inequality holds
    \begin{equation}
    \label{8}
    \left(G^\perp M_dM^{-1}\frac{\partial (G^\perp \nabla V)}
    {\partial q}\right)_{q=q^*}>0.
    \end{equation}
    \carrew
\end{proposition}
\vspace{2mm} \textbf{Proof}:
Differentiate both side of PDE
(\ref{7}) respect to $q$:
    \begin{equation}
    \label{9}
    \frac{\partial (G^\perp \nabla V)}{\partial q}=\Big(G^\perp M_dM^{-1}\frac{\partial^2 V_d}{\partial q^2}\Big)^T+\frac{\partial G^\perp M_dM^{-1}}{\partial q}\nabla V_d.
    \end{equation}
Note that $\left.\nabla V_d\right|_{q=q^*}=0_n$. Thus, (\ref{9}) at $q=q^*$ is
    \begin{equation*}
    \left.\frac{\partial (G^\perp \nabla V)}{\partial q}\right|_{q=q^*}=\left.\Big(G^\perp M_dM^{-1}\frac{\partial^2 V_d}{\partial q^2}\Big)^T\right|_{q=q^*}=\Big(\left.\frac{\partial^2 V_d}{\partial q^2}(G^\perp M_dM^{-1})^T\Big)\right|_{q=q^*}
    \end{equation*}
To complete the proof, multiply both side of above equation from
left to $\left.\big(G^\perp M_dM^{-1}\big)\right|_{q=q^*}$ and
notice that arbitrary matrix $A$ is positive definite if
$\xi^TA\xi>0$ for any $\xi\neq 0$.
    \carre
As explained before, we suppose that $m=n-1$. Thus, with a minor
loss of generality, suppose that \jgrv{$G=P[I_m,0_{m\times n-m}]^T$ with $P$ a permutation matrix which results in $G^\bot=e_k^T,k\in\bar{n}$}. Simplify PDE
(\ref{6}) term by term as follows. The first term is:
        \begin{equation*}
\label{11}
G^\bot (q)\nabla_q \big(p^TM^{-1}(q)p\big)=p^T\frac{\partial M^{-1}}{\partial q^{(k)}}p
\end{equation*}
In the sequel, the following notations are used
        \begin{equation}
\label{12}
M^{-1}=\frac{1}{\det{M}} \mathfrak{M}(q) \quad \implies \quad \frac{\partial M^{-1}}{\partial q^{(k)}}=\frac{1}{(\det{M})^2}\mathbb{M}(q)
\end{equation}
where $\mathfrak{M}\in \mathbb{R}^{n\times n}$ is adjugate matrix of
$M$ and $\mathbb{M}\in \mathbb{R}^{n\times n}$ is matrix of
nominator elements  of $\frac{\partial M^{-1}}{\partial q^{(k)}}$.
Note that in spite of most previously reported research on this
topic, it is not assumed that $M(q)$ merely depends on some
specified configuration variables. \jgrv{Regard to second term of (\ref{6}),} it is assumed that
$M_d^{-1}(q)$ has the following structure:
        \begin{equation}
\label{13}
M_d^{-1}=
\begin{bmatrix}
a_1 & 0 & \dots & b_1 & 0 & \dots & 0 \\
0 & a_2  & \dots & b_2 & 0 & \dots & 0 \\
\vdots & \vdots & & \vdots &  &  & 0 \\
b_1 & b_2 & \dots & a(q) & b_k & \dots & b_{n-1} \\
0 & 0 & \dots & b_k & a_k & \dots & 0 \\
\vdots & \vdots & & \vdots &  & \ddots & \vdots\\
0 & 0 & \dots & b_{n-1} & 0 & \dots & a_{n-1}
\end{bmatrix}
\end{equation}
in which, all the elements of it are zero except diagonal elements,
and the $k$-th row and column. Notice that $a(q)$ is the only
element which is state dependent. In other words, since our aim is
to solve PDE of kinetic energy as simple as possible or at least
simplify it, $a_i$s and $b_i$s are considered to be constant. Notice
that the most important property of this structure is that $k$-th
row of adjugate matrix $\mathfrak{M}$ is independent of
configuration variables and $b_i$s.

In order to simplify second term of (\ref{6}),  $G^\bot M_dM^{-1}$
is represented as follows
        \begin{equation}
\label{14}
G^\bot M_dM^{-1}=\frac{1}{\det{M}\det{M_d^{-1}}}\gamma
\end{equation}
where $\gamma\in \mathbb{R}^n$ is  a row vector independent of
$a(q)$. This is another advantage of the selected form of
(\ref{13}). determinant of ${M_d^{-1}}$ is
\begin{equation*}
 \det{M_d^{-1}}=\phi_1a(q)+\phi_2,
 \end{equation*}
where $\phi_1,\phi_2$ are constant parameters depending on other
elements of $M_d^{-1}$. Finally, second term of (\ref{6}) may be
reduced to
        \begin{equation}
\label{15}
G^\bot M_dM^{-1}\nabla_q \big(p^TM_d^{-1}(q)p\big)=\frac{p^T\displaystyle\sum_{i=1}^{n}\bigg(\gamma^{(i)}\frac{\partial M_d^{-1}}{\partial q^{(i)}}\bigg)\hspace{1mm}p}{\det{M}\det{M_d^{-1}}}
\end{equation}
in which all elements of $\frac{\partial M_d^{-1}}{\partial
q^{(i)}}$ are zero except the $(k,k)$ element. Notice that if $M_d$
was selected like what is reported in previous
works~\cite{acosta2005interconnection, gomez2001stabilization} as a
function of only $q^{(k)}$, then the above equation reduces to
\begin{equation*}
\frac{\gamma^{(k)}p^T\frac{\partial M_d^{-1}}{\partial
q^{(k)}}p}{\det{M}\det{M_d^{-1}}}.
\end{equation*}
In order to simplify the last term of (\ref{6}), as reported in
\cite{acosta2005interconnection}, $J_2$ is linear with
respect to $p$. Therefore, $J_2$ can be parameterized in the
following form
\begin{equation}
    \label{17}
        J_2(q,p)=
\frac{1}{\det{M}}\begin{bmatrix}
    0 & {p}^T\alpha_1(q) & \dots & {p}^T\alpha_{n-1}(q) \\
    {p}^T\alpha_n(q) & 0 & \dots & {p}^T\alpha_{2n-2}(q) \\
    \vdots & \vdots & \ddots & \vdots \\
    {p}^T\alpha_{n^2-2n+2}(q) & {p}^T\alpha_{n^2-2n+3}(q) & \dots & 0
    \end{bmatrix}
    \end{equation}
where  $\alpha_i\in \mathbb{R}^n, i\in \overline{n(n-1)}$. Note that
this form of $J_2$ is not generally  skew-symmetric. However, only a
row of this matrix will be determined, thus, the column
corresponding to this row will be selected in such a way that $J_2$
becomes skew-symmetric. Other elements of this matrix are free
design parameters. Invoking \cite{acosta2005interconnection}, $J_2$
may be rewritten  as follows
\begin{equation*}
J_2=\frac{1}{\det{M}}\displaystyle\sum_{i=1}^{n_0} {p}^T\alpha_i W_i, \qquad n_0=n(n-1),
\end{equation*}
where  $W_i$ are  matrices which are set as follows
\begin{equation*}
\begin{split}
W_1=&W^{1,2}, W_2=W^{1,3}, \dots, W_{n-1}=W^{1,n}, W_n=W^{2,1}, \dots, W_{n_{0}}=W^{n,n-1},
\end{split}
\end{equation*}
in which $W^{i,j}$ is a matrix such that all of its elements are
zero except the $(i,j)$ element which is equal to 1. Hence,
$G^{\perp}J_2$ can be written as follows
\begin{equation*}
G^{\perp}(q)J_2(p,q)=\frac{1}{\det{M}}{p}^T\mathcal{J}(q)A,\quad \mathcal{J}=
\begin{bmatrix}
\alpha_1 & \dots & \alpha_{n_0}
\end{bmatrix}
\in R^{n \times n_0},\quad A=
\begin{bmatrix}
(G^{\perp}W_1)^T & \dots & (G^{\perp}W_{n_0})^T
\end{bmatrix}^T
\in R^{n_0 \times n}.
\end{equation*}
Thus, $\mathcal{J}A$ may be written as:
\begin{equation*}
\begin{split}
\mathcal{J}A&=[\alpha_{(k-1)n-k+2},\dots,\alpha_{(k-1)n},0_n,\alpha_{(k-1)n+1},\dots,\alpha_{kn-k}]\triangleq B(q)\in\mathbb{R}^{n\times n}.
\end{split}
\end{equation*}
Notice that just one of the rows of $J_2$ appears in this equation. Finally,
 third term in PDE (\ref{6}) is reduced to
\begin{equation}
\label{23}
G^\bot J_2(q,p)M_d^{-1}(q)p=\frac{1}{\det{M}}p^TBM_d^{-1}p
\end{equation}
All of the terms in (\ref{6}) are quadratic with respect to $p$ and
should be symmetric. Replacing (\ref{12}), (\ref{15}) and (\ref{23})
in (\ref{6}) results in the following relation:
 \begin{equation}
\label{24}
\frac{\mathbb{M}}{\det{M}}-\frac{\displaystyle\sum_{i=1}^{n}\gamma^{(i)}\frac{\partial M_d^{-1}}{\partial q^{(i)}}}{\det{M_d^{-1}}}+(BM_d^{-1}+M_d^{-1}B^T)=0
\end{equation}
There are $\frac{n(n+1)}{2}$ equations  and $n(n-1)$   free
parameters in  above relation. At first, it seems that for $n\geq 3$
there is no need to calculate
$\displaystyle\sum_{i=1}^{n}\gamma^{(i)}\frac{\partial
M_d^{-1}}{\partial q^{(i)}}$. However, invoking Lemma2 in
\cite{acosta2005interconnection}, it is easy to show that rank of
$(BM_d^{-1}+M_d^{-1}B^T)$ is always $n-1$. It is also shown in
\cite{crasta2015matching} that the number of PDEs which should be
solved is $\frac{1}{6}s(s+1)(s+2)$. Therefore, equality (\ref{24})
leads to  $\frac{n(n+1)}{2}-1$ algebraic equations and one PDE with
respect to $a(q)$. Notice that base on the structure of $M_d^{-1}$,
all the elements of the second term are zero except the $(k,k)$
element. Therefore,  the $(k,k)$ element of equation (\ref{24})
leads to a PDE and other elements results in simple algebraic
equations. These algebraic equations are derived by some
manipulation as follows:
 \begin{align}
\label{25}
&\frac{1}{\det{M}}
    [\mathbb{M}^{(11)}, \mathbb{M}^{(12)}, \dots, \mathbb{M}^{(1n)}, \mathbb{M}^{(22)},\dots, \mathbb{M}^{((k-1)k)}, \mathbb{M}^{(k(k+1))}, \dots, \mathbb{M}^{(nn)}]^T\nonumber\\&= -\Psi
    [\alpha_{(k-1)n-k+2}^{(1)},\alpha_{(k-1)n-k+2}^{(2)} \dots, \alpha_{(k-1)n-k+2}^{(n)}, \alpha_{(k-1)n-k+3}^{(1)}, \dots,\alpha_{(k-1)n-k+3}^{(n)},\dots, \alpha_{kn-k}^{(1)}, \dots \alpha_{kn-k}^{(n)}]^T
\end{align}
where
 \begin{equation*}
\Psi=\begin{bmatrix}
\psi_1 & \dots & \psi_{n(n-1)}
\end{bmatrix}
\in \mathbb{R}^{\frac{n(n+1)-2}{2}\times n(n-1)}
\end{equation*}
 \begin{equation*}
\begin{split}
&\psi_1=[2a_1, 0_{k-2}^T, b_1, 0_{\frac{n(n+1)}{2}-k-1}^T]^T,\qquad
\psi_2=[0, a_1, 0_{n+k-4}^T, b_1, 0_{\frac{n(n+1)}{2}-n-k}^T]^T,\qquad \dots  \\
&\psi_{n+1}=[0, a_2,0_{k-3}^T, b_1, 0_{\frac{n(n+1)}{2}-k-1}^T]^T, \quad\dots\quad
\psi_{n(n-1)}=[0_{\frac{k(2n-k-1)}{2}-1}^T, b_{n-1},0_{\frac{n(n+1)-k(2n-k-1)}{2}-1}^T]^T,
\end{split}
\end{equation*}
in which $\alpha_i^{(j)}$ is the $j$-th element of vector
$\alpha_i$. Matrix $\Psi$ is generally full rank; therefore,
equation (\ref{25}) has at least one solution. The remaining PDE is
given by
 \begin{equation}
\label{27}
\frac{\displaystyle\sum_{i=1}^{n}\gamma^{(i)}\frac{\partial a(q)}{\partial q^{(i)}}}{\det{M_d^{-1}}}-\frac{\mathbb{M}^{(kk)}}{\det{M}}-2\displaystyle\sum_{i=1}^{n-1} b_i\alpha_{(k-1)n-k+1+i}^{(k)}=0.
\end{equation}
Note that $a_i$s and $b_i$s should be determined such that $\Psi$ is full rank, $M_d^{-1}$ is positive definite and proposition \ref{pr1} is satisfied.

The following statements can be verified for (\ref{24}):
\begin{itemize}
    \item If the assumption of \cite{gomez2001stabilization} holds, i.e. $M$ is
    only function of unactuated coordinate $q^{(k)}$, the second term of (\ref{24})
     is reduced to $\frac{\gamma^{(k)}\frac{\partial M_d^{-1}}{\partial q^{(k)}}}
     {\det M_d^{-1}}$ and PDE (\ref{27}) is replaced by the following ODE
    \begin{equation}
    \label{28}
    \begin{split}
    &\frac{\gamma^{(k)}\frac{d a(q^{(k)})}{d q^{(k)}}}{\phi_1a(q^{(k)})+\phi_2}=\frac{\mathbb{M}^{(kk)}}{\det{M}}+2\displaystyle\sum_{i=1}^{k-1} b_i\alpha_{(k-1)n-k+1+i}^{(k)}=f(q^{(k)})
    \end{split}
    \end{equation}
Analytic solution of this ODE is
    \begin{equation}
\label{29}
\begin{split}
a(q^{(k)})=\frac{\lambda e^{\phi_1F(q^{(k)})}-\phi_2}{\phi_1}
\end{split}
\end{equation}
where $\lambda$ is a constant parameter and $F=\int
\frac{f}{\gamma^{(k)}} dq^{(k)}$. Note that in the method proposed
in \cite{gomez2001stabilization}, the obtained ODEs is generally a
set of ODE and has analytic solution if $n=2$.

\item \label{it2} If $\mathbb{M}^{(kk)}$ is equal to zero, then the second term
in (\ref{27}) is omitted. Hence, it is easy to select $a(q)$ with respect to free
 parameters. Special cases of this condition is considered in
 \cite{acosta2005interconnection} where an analytic solution is proposed in
 which $a(q)$ is merely \jgrv{a function of one of the} $q^{(i)}$s. In the proposed solution,
 since the aim is to derive a simple solution, $M_d$ will be considered to be a
 constant matrix such that Proposition~\ref{pr1} is satisfied.
 Note that if
  $M$ is constant, we can choose a constant value for $M_d$ without considering
  a special form for $G(q)$.  The proposed IDA-PBC in \cite{d2006further} for acrobot is an example of this case.

\item  In other cases, a PDE should be solved. One may invoke the methods
proposed in~\cite{acosta2009pdes,viola2007total} to simplify it.
Another powerful method is using Pfaffian differential equations
detailed in~\cite{harandi2020solution,sneddon2006elements}. Base on this method, the
corresponding Pfaffian equations to PDE (\ref{27}) are
\begin{equation*}
\frac{\det {M_d^{-1}}dq^{(1)}}{\gamma^{(1)}}=\dots=\frac{\det
{M_d^{-1}}dq^n}{\gamma^{(n)}}=\frac{d
a}{\frac{\mathbb{M}^{(kk)}}{\det{M}}+2\displaystyle\sum_{i=1}^{n-1}
b_i\alpha_{(k-1)n-k+1+i}^{(k)}}
\end{equation*}
In \cite{harandi2020solution} (see also
\cite[ch.2]{sneddon2006elements}) some tips are proposed to solve
Pfaffian differential equations.
\end{itemize}
In the next section, some illustrative case studies are examined to
show the applicability of proposed method. Note that similar to
\cite{viola2007total,donaire2016simultaneous} we only concentrate on the matching equation
related to kinetic energy.

\section{Case Studies}\label{s4}
In the following three case studies is proposed to verify the three
above mentioned statements. The first example is Pendubot in which
its matching equation is replaced by an ODE. The second example is
VTOL aircraft where the corresponding PDE is solved easily by a
constant $M_d$. The last example is 2D SpiderCrane in which its
matching equation is solved by Pffafian differential equations.
\subsection{Pendubot} \label{pen}
Pendubot is a 2R planar serial robot in which the first joint is
only actuated. In~\cite{sandoval2008interconnection} an IDA-PBC
controller is designed for this robot by suitably defining new
variables. In here the PDE of kinetic energy shaping is solved
systematically. Inertia matrix and potential energy of this robot
are given by:
\begin{equation}
\label{45}
M=
\begin{bmatrix}
c_1+c_2+2c_3\cos(q^{(2)}) & c_2+c_3\cos(q^{(2)}) \\ c_2+c_3\cos(q^{(2)}) & c_2
\end{bmatrix},\qquad
V=c_4g\cos(q^{(1)})+c_5g\cos(q^{(1)}+q^{(2)}),\qquad G=\begin{bmatrix}
1 \\ 0
\end{bmatrix},
\end{equation}
with $c_i$s defined in \cite{sandoval2008interconnection}. After
some manipulation, the following expressions are obtained
\begin{equation}
\label{46}
\begin{split}
&\det{M}=c_1c_2-c_3^2\cos^2(q^{(2)}), \qquad \mathfrak{M}=
\begin{bmatrix}
c_2 & -c_2-c_3\cos(q^{(2)}) \\ -c_2-c_3\cos(q^{(2)}) & c_1+c_2+2c_3\cos(q^{(2)})
\end{bmatrix},\\
&\mathbb{M}=
\begin{bmatrix}
-2c_3^2\sin(q^{(2)})\cos^2(q^{(2)}) & \begin{array}{c} c_1c_2c_3\sin(q^{(2)})+c_3^3\sin(q^{(2)})\cos^2(q^{(2)})\\+2c_2c_3^2\sin(q^{(2)})\cos(q^{(2)})\end{array} \\ \begin{array}{c} c_1c_2c_3\sin(q^{(2)})+c_3^3\sin(q^{(2)})\cos^2(q^{(2)})\\+2c_2c_3^2\sin(q^{(2)})\cos(q^{(2)})\end{array}  & \begin{array}{c} -2c_1c_2c_3\sin(q^{(2)})-2c_3^3\sin(q^{(2)})\cos^2(q^{(2)})\\-2c_3^2(c_1+c_2)\sin(q^{(2)})\cos(q^{(2)})  \end{array}
\end{bmatrix},\\
&M_d^{-1}=
\begin{bmatrix}
a_1 & b_1 \\ b_1 & a(q^{(2)})
\end{bmatrix},\quad J_2=
\begin{bmatrix}
0 & p^T\alpha_1 \\ p^T\alpha_2 & 0
\end{bmatrix},\quad
\gamma^T=\begin{bmatrix}-c_2b_1-c_2a_1-c_3a_1\cos(q^{(2)}) \\ a_1c_1+a_1c_2+b_1c_2+b_1c_3\cos(q^{(2)})+2a_1c_3\cos(q^{(2)})  \end{bmatrix}
\end{split}
\end{equation}
Note that in the following $\alpha_2$ will be determined and $\alpha_1=-\alpha_2$ will be set.
Equation (\ref{24}) for this case is derived as follows
\begin{equation}
\label{47}
\frac{\gamma^{(2)}}{\det{M_d^{-1}}}
\begin{bmatrix}
0 & 0 \\ 0 & \frac{\partial a}{\partial q^{(2)}}
\end{bmatrix}
=\frac{1}{\det{M}}\mathbb{M}+
\begin{bmatrix}
2\alpha_{2}^{(1)}a_1 & \alpha_{2}^{(2)}a_1+\alpha_{2}^{(1)}b_1 \\ \alpha_{2}^{(2)}a_1+\alpha_{2}^{(1)}b_1 & 2\alpha_{2}^{(2)}b_1
\end{bmatrix}
\end{equation}
By solving two algebraic equations, $\alpha_2$ is obtained as follows
\begin{equation}
\label{48}
\begin{split}
&\alpha_{2}^{(1)}=-\frac{\mathbb{M}^{(11)} }{2a_1\det{M}}=\frac{c_3^2\sin(q^{(2)})\cos^2(q^{(2)})}{a_1\big(c_1c_2-c_3^2\cos^2(q^{(2)})\big)},\\
&\alpha_{2}^{(2)}=-\frac{\mathbb{M}^{(21)} }{a_1\det{M}}-\frac{\alpha_{2}^{(1)}b_1}{a_1}=-\frac{c_1c_2c_3\sin(q^{(2)})+c_3^3\sin(q^{(2)})\cos^2(q^{(2)})+2c_2c_3^2\sin(q^{(2)})\cos(q^{(2)})}{a_1\big(c_1c_2-c_3^2\cos^2(q^{(2)})\big)}\\&-\frac{b_1c_3^2\sin(q^{(2)})\cos^2(q^{(2)})}{a_1^2\big(c_1c_2-c_3^2\cos^2(q^{(2)})\big)}
\end{split}
\end{equation}
Finally, the following ODE should be solved
\begin{equation}
\label{49}
\frac{1}{a_1a(q^{(2)})-b_1^2}\frac{da}{dq^{(2)}}=\frac{1}{\gamma^{(2)}\det{M}}\bigg(\mathbb{M}^{(22)}-\frac{-2b_1\mathbb{M}^{(21)}}{a_1}+\frac{b_1^2\mathbb{M}^{(11)}}{a_1^2}\bigg)
\end{equation}
This ODE is in the form of (\ref{28}) and its solution is derived from (\ref{29}) with
\begin{equation*}
\phi_1=b_1,\quad \phi_2=-b_1^2,\quad F(q^{(2)})=\int \frac{1}{\gamma^{(2)}\det{M}}\bigg(\mathbb{M}^{(22)}-\frac{-2b_1\mathbb{M}^{(21)}}{a_1}+\frac{b_1^2\mathbb{M}^{(11)}}{a_1^2}\bigg) dq^{(2)}.
\end{equation*}
For example, assume that $c_1=4, c_2=1$ and $c_3=1.5 $. By some
manipulation, $a(q^{(2)})$ is obtained  as follows
\begin{equation}
\label{52}
a(q^{(2)})=\cos(q^{(2)})^{-7/3}+(4-3\cos(q^{(2)}))^{49/6}-(4+3\cos(q^{(2)}))^{-7/2},
\end{equation}
where  $a_1=1,b_1=-5,\lambda=1$ are chosen to simplify the ODE (\ref{49}) and also the necessary condition (\ref{8}) is satisfied. The solution of potential energy PDE is proposed in Appendix.

\subsection{VTOL Aircraft}
Dynamic model of VTOL in PCH form (\ref{1}) is given as follows
\begin{equation}
\label{26}
G(q)=
\begin{bmatrix}
-\sin(\theta) & \epsilon\cos(\theta) \\
\cos(\theta) & \epsilon\sin(\theta) \\
0 & 1
\end{bmatrix},\qquad\qquad
M=I, \qquad \qquad
V=gy,\qquad \qquad q=\begin{bmatrix}
x\\y\\ \theta
\end{bmatrix}
\end{equation}
where, $x$ and $y$ denote the position of center of mass, $\theta$
is the roll angle and $\epsilon$ models  the effect of the slopped
wings. The desired equilibrium point of the system is
$[x_*,y_*,0]^T$. In \cite{acosta2005interconnection} a controller
with state-dependent $M_d$ by defining new inputs is derived. Since the
inertia matrix is constant, it is possible to solve the PDE of
kinetic energy with a constant $M_d$, represented by:
\begin{equation*}
M_d=
\begin{bmatrix}
a & d & e \\ d & b & f \\ e & f & c
\end{bmatrix}
\end{equation*}
Necessary condition (\ref{8}) in this case leads to following
inequality
\begin{equation*}
\left.\Big(g\cos(\theta)(\epsilon\cos(\theta)+f\sin(\theta)-c\epsilon)\Big)\right|_{\theta=0}>0.
\end{equation*}
A suitable choice for the matrix parameters is
\begin{equation*}
a=\kappa\epsilon^2,\quad b=1,\quad c=\kappa',\quad d=0,\quad e=\epsilon,\quad f=0,
\end{equation*}
where the constants $\kappa,\kappa'>0$ should be selected such that
$\kappa\kappa'>1$. Note that $M_d=I$ does not satisfy the necessary
condition (\ref{8}) which is in line with our prior knowledge that
it is not possible to stabilize the system with merely potential
energy shaping. Although solving the potential energy PDE (\ref{7})
is out of scope of this paper, but in this case its solution with
$\kappa=20$ and $\kappa'=0.1$ is derived as follows
\begin{equation*}
\begin{split}
&V_d=\Big(\epsilon(y-y^*)+\ln\big(\epsilon\cos(\theta)-0.1\epsilon\big)\Big)^2+\Big(\frac{1}{20\epsilon}(x-x^*)-(\theta-\theta^*)-0.1\text{arctanh}\big(1.1055\tan(\frac{\theta}{2})\big)\Big)^2\\&-2\epsilon\ln(0.9\epsilon)(y-y^*)-\frac{g-2\epsilon\ln(0.9\epsilon)}{g\epsilon}\ln\big(\epsilon\cos(\theta)-0.1\epsilon\big).
\end{split}
\end{equation*}
\begin{figure}[b]
    \centering
    \subfigure[The stabilization errors and control effots with proposed controller.]{
        \includegraphics[width=.45\linewidth]{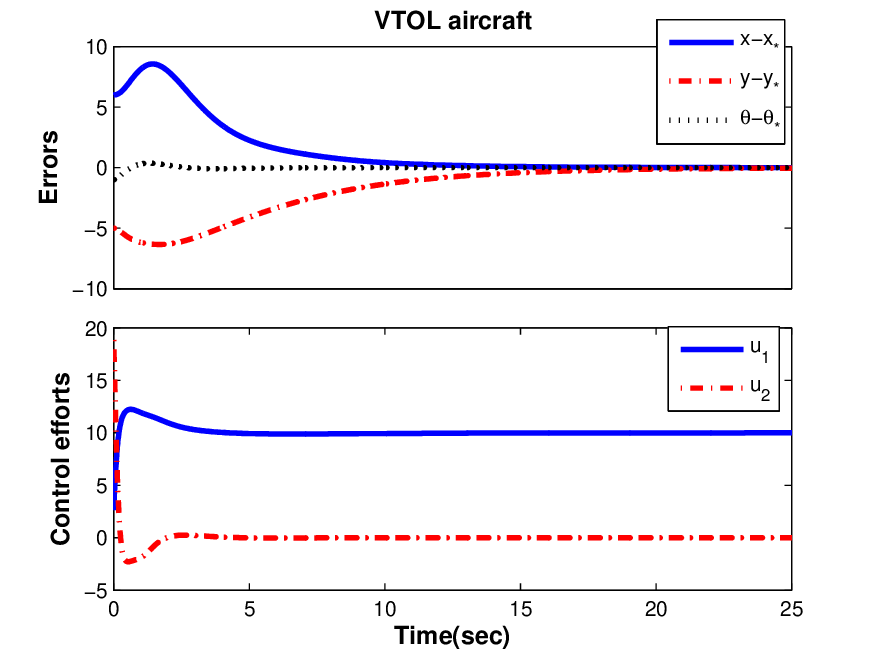}
        \label{p31}
    }\hspace{1mm}
    \subfigure[The motion of VTOL aircraft in plene.]{
        \includegraphics[width=.45\linewidth]{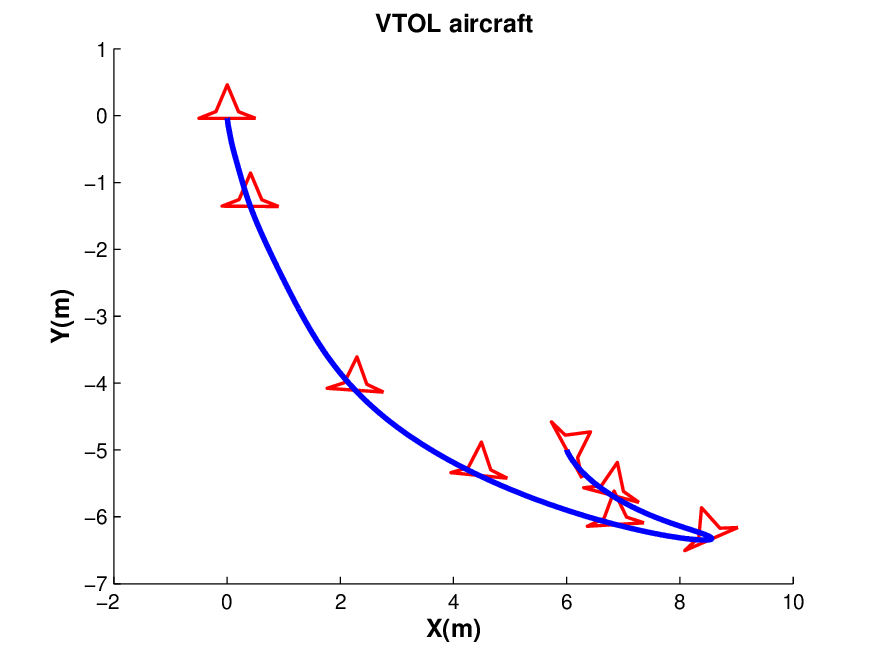}
    }
        \label{p32}
    \caption{Simulation results of proposed controller on VTOL aircraft. The aircraft moves toward its desired position with smooth states and control law.}
    \label{p3}
\end{figure}
\begin{wrapfigure}{r}{0.5\textwidth}
    \vspace{-1pt}
    \begin{center}\includegraphics[scale=0.4]{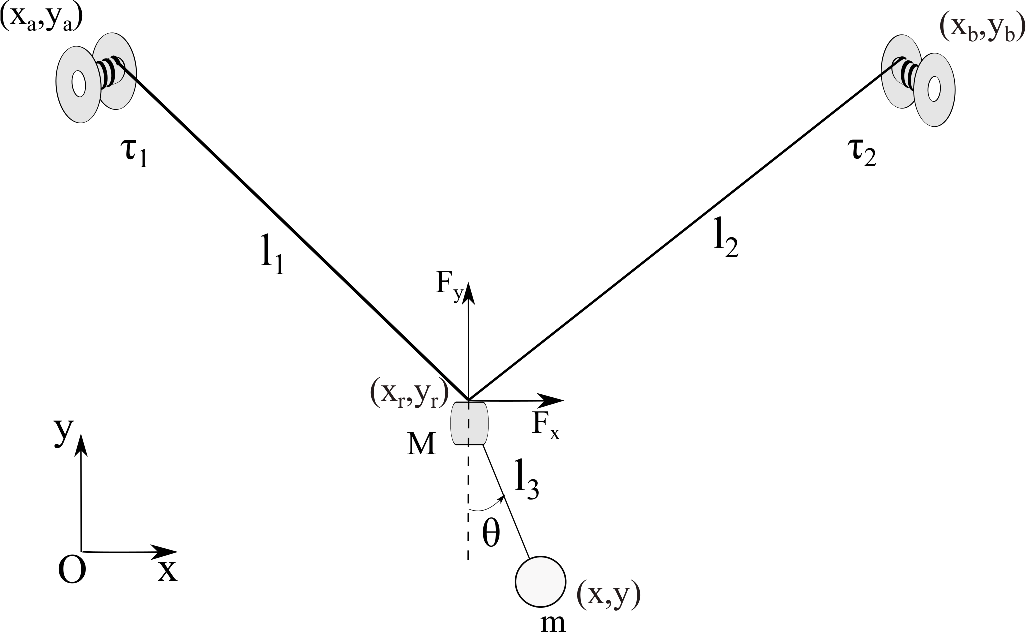}
    \end{center}
    \vspace{-15pt}
    \caption{\label{fig:Sch} Schematic of 2D SpiderCrane system.}
    \vspace{-8pt}
\end{wrapfigure}
To simulate the response, consider the initial condition as
$q(0)=[6,-5,-1]^T$ with zero velocity while the desired position is
$q_*=[0,0,0]^T$. Set, $\epsilon=0.3$ and $K_v=\mbox{diag}\{1,0.5\}$.
Simulation results are illustrated in Fig.~\ref{p3}. As shown in
Fig.~\ref{p31}, the errors converge to zero in about 20 second with
an acceptable control efforts amplitude. The motion of the robot in
$X-Y$ plane is depicted in Fig.~\ref{p31}. Due to coupling of
inputs, the aircraft first moves to farther position to correct its
orientation and then goes to the desired position. Note that the
advantage of the proposed controller in comparison with that
reported in \cite{acosta2005interconnection} is its simplicity.

\subsection{2D SpiderCrane}
This system consists of a load suspended from a ring which is
controlled by two cables. The schematic of this system is
depicted in Fig.~\ref{fig:Sch}. The position of the ring and the
mass are denoted by $(x_r,y_r)$ and $(x,y)$, respectively, and their
mass is denoted by $M$ and $m$, respectively. The length of the
controlled cables is denoted by $l_1$ and $l_2$, while $l_3$ denotes
the fixed length of the cable between ring and the mass. Dynamic
equation of the system is in the form (\ref{1}) with following
parameters
\begin{equation*}
\hspace{-2mm}
q=\begin{bmatrix}
x_r \\ y_r\\ \theta
\end{bmatrix},\quad
G=\begin{bmatrix}
1 & 0 \\ 0 & 1 \\ 0 & 0
\end{bmatrix}^T,\quad
V=(M+m)gy_r-mgl_3\cos(\theta), \quad
M(q)=\begin{bmatrix}
M+m & 0 & ml_3\cos(\theta) \\
0 & M+m & ml_3\sin(\theta) \\
ml_3\cos(\theta) & ml_3\sin(\theta) & ml_3^2
\end{bmatrix}.
\end{equation*}
Two IDA-PBC controller have been designed for SpiderCrane. In
\cite{kazi2008stabilization} merely the potential energy is shaped
while in \cite{sarras2010total} total energy shaping method proposed
in  \cite{acosta2005interconnection} is used such that first a
partial feedback linearization is applied to the system and then a
desired inertia matrix which is merely a function of $\theta$ is
chosen. In here, the aim is to derive a more general solution such
that $M_d$ may be set as a function of $x_r$ and $y_y$. Consider
$M_d^{-1}$ in the form of (\ref{13}). One can easily check that
necessary condition (\ref{8}) is satisfied if $b_1ml_3+a_2(M+m)>0$.
In order to solve matching equation (\ref{6}), the following
parameters are derived
\begin{equation*}
\begin{split}
&\det M(q)=(M+m)^2ml_3^2-(M+m)m^2l_3^2,\quad  \\&\mathfrak{M}=\begin{bmatrix}
(M+m)ml_3^2-m^2l_3^2\sin^2(\theta) & m^2l_3^2\sin(\theta)\cos(\theta) & -(m+M)ml_3\cos(\theta) \\ m^2l_3^2\sin(\theta)\cos(\theta) & (M+m)ml_3^2-m^2l_3^2\cos^2(\theta) & -(M+m)ml\sin(\theta) \\ -(m+M)ml_3\cos(\theta) & -(M+m)ml\sin(\theta) & (M+m)^2
\end{bmatrix},
\\&\mathbb{M}=\det M(q)\begin{bmatrix}
-2m^2l_3^2\sin(\theta)\cos(\theta) & m^2l_3^2\cos(2\theta) & (M+m)ml_3\sin(\theta) \\ m^2l_3^2\cos(2\theta) & m^2l^2\sin(2\theta) & -(M+m)ml_3\cos(\theta) \\ (M+m)ml_3\sin(\theta) & -(M+m)ml_3\cos(\theta) & 0
\end{bmatrix},\\&
\gamma^T=\begin{bmatrix}
-a_2b_1(M+m)ml_3^2+a_2b_1m^2l_3^2\sin^2(\theta)-a_1b_2m^2l_3^2\sin(\theta)\cos(\theta)-a_1a_2(M+m)ml_3\cos(\theta) \\ -a_2b_1m^2l_3^2\sin(\theta)\cos(\theta)-a_1b_2(M+m)ml_3^2+a_1b_2m^2l_3^2\cos^2(\theta)-a_1a_2(M+m)ml_3\sin(\theta) \\ a_2b_1(M+m)ml_3\cos(\theta)+a_1b_2(M+m)ml_3\sin(\theta)+a_1a_2(M+m)^2
\end{bmatrix}
\end{split}
\end{equation*}
Based on necessary condition and simplifying the corresponding
matching equation, we choose $b_1=b_2=0$. By this means, the matrix
$\Psi$ in equality (\ref{25}) is in the following form
\begin{equation*}
\Psi=\begin{bmatrix}
2a_1 & 0 & 0 & 0 & 0 & 0 \\
0 & a_1 & 0& a_2& 0& 0 \\
0 & 0 & a_1 & 0 & 0 & 0 \\
0 & 0 & 0 & 0 & 2a_2 & 0 \\
0 & 0 & 0 & 0 & 0 & a_2
\end{bmatrix}
\end{equation*}
This matrix is full rank, hence $[\alpha_5^T,\alpha_6^T]^T$ is
determined by right pseudo-inverse of $\Psi$. The Pfaffian
differential equations of PDE (\ref{27}) for this system is given as
follows
\begin{equation*}
\frac{dx}{-ml_3\cos(\theta)}=\frac{dy}{-ml_3\sin(\theta)}=\frac{d\theta}{M+m}=\frac{d a}{0}
\end{equation*}
The solutions to these equations are
\begin{equation*}
x+\frac{ml_3}{M+m}\sin(\theta)=c_1,\qquad\qquad y-\frac{ml_3}{M+m}\cos(\theta)=c_2,
\end{equation*}
with $c_1$ and $c_2$ as free parameters.
Invoking \cite{harandi2020solution} $a(q)$ is
\begin{equation*}
a(q)=\phi\Big(x+\frac{ml_3}{M+m}\sin(\theta),y-\frac{ml_3}{M+m}\cos(\theta)\Big),
\end{equation*}
where $\phi$ is an arbitrary function. General form of $V_d$ is proposed in Appendix.

\section{Conclusions and Future Prospects}\label{s5}
In this paper a systematic method to simplify the matching equation
related to kinetic energy shaping for underactuated robots with one
degree of underactuation was proposed. A special structure of
desired inertia matrix was considered in such a way that just one of
its elements depends on configuration variables. By this means, the
arisen PDE can be analytically solved for robots with \jgrv{some properties including
manipulators} with inertia matrix depending on just one variable. The
proposed method was successfully implemented on VTOL aircraft,
pendubot and 2D SpiderCrane. Extension of this method to robots with
more degrees of underactuation and also consideration of potential
energy PDE are currently being examined in our research group.

\bibliographystyle{model1-num-names}
\bibliography{ref}
\end{document}